\pdfoutput=1
\RequirePackage{fix-cm}
\documentclass[smallextended]{svjour3}       
\smartqed  
\usepackage{graphicx}
 \journalname{submitted}

\newcommand{\rmi}{\mathrm{i}}

\newcommand{\RMP}{Rev. Mod. Phys.}
\newcommand{\PC}{Physica C }

\newcommand{\JSNM}{J. Supercond. Nov. Magn. }
\newcommand{\PRB}{Phys. Rev. B }
\newcommand{\PRL}{Phys. Rev. Lett. }
\newcommand{\APL}{Appl. Phys. Lett. }
\newcommand{\JAP}{J. Appl. Phys. }
\newcommand{\SUST}{Supercond. Sci. Technol. }

\begin{document}
\title{Superconducting and structural properties of Nb/PdNi/Nb trilayers.}

\titlerunning{Superconducting and structural properties of Nb/PdNi/Nb trilayers.}        

\author{N. Pompeo 	\and
        K. Torokhtii	\and
        C. Meneghini	\and
	    S. Mobilio		\and
	    R. Loria		\and
        C. Cirillo		\and
        E. A. Ilyina	\and
        C. Attanasio	\and
        S. Sarti		\and
        E. Silva
}

\institute{N. Pompeo  (\email{pompeo@fis.uniroma3.it})
            \and
            K. Torokhtii
            \and
            R. Loria
            \and
            E. Silva  (\email{silva@fis.uniroma3.it})
	    \at
            Dipartimento di Fisica ``E. Amaldi'' and Unit\`a CNISM, Universit\`a Roma Tre, Via della Vasca Navale 84, 00146 Roma, Italy  
            \and
	        C. Meneghini
            \and
            S. Mobilio
\at            Dipartimento di Fisica ``E. Amaldi'' Universit\`a Roma Tre, Via della Vasca Navale 84, 00146 Roma, Italy and CNR-OGG Grenoble, c/o ESRF (Grenoble, France)
	    \and
	    C. Cirillo
	    \and
	    E. A. Ilyina
	    \and
	    C. Attanasio
	    \at
	    CNR-SPIN and Dipartimento di Fisica ``E. R. Caianiello", Universit\`a di Salerno, 84084 Fisciano (SA), Italy
	    \and
	    S. Sarti
	    \at
	    Dipartimento di Fisica, Universit\`a ``La Sapienza", 00185 Roma, Italy
}

\date{May 9th, 2012}

\maketitle

\begin{abstract}

The superconducting and structural properties of S/F/S (Superconductor/Ferromagnet/Superconductor) heterostructures have been studied by means of microwave measurements (1-20 GHz) and x-ray absorption fine structure (XAFS) spectroscopy. Nb/PdNi/Nb trilayers have been studied as a function of F layer thicness. With respect to pure Nb, XAFS analysis shows that the heterostructures exhibit larger structural disorder in the S layers. Microwave measurements show evidence for a progressively weaker vortex pinning with increasing F thickness. However, no clear correlation is found with the local disorder in Nb: the weakest pinning is not in the most disordered trilayer. Therefore the structural disorder in the superconducting material cannot explain on its own the changes in vortex pinning. We argue that the F layer acts on the superconducting state itself. We propose possible explanations for the observed behavior.

\end{abstract}

\section{Introduction}
\label{sec:intro}

The physics of Superconducting/Ferromagnetic (S/F) heterostructures allows to investigate the competing magnetic (F) and superconducting (S) ordering in artificial structures where a spatial separation among S and F exists.
The interest in such systems has been revamped in recent years \cite{buzdinRMP05,lyuksyutovAdP05}, in particular with respect to the many effects that can arise when the ferromagnetic coherence length is of the order of the F layer thickness. Similarly, a new interest in the vortex physics (namely, vortex pinning) in S/F structures has arosen  \cite{silhanekAPL06,blamirePRL07}.

However, measurements of the vortex state response at high frequencies (radio and microwave) are lacking, despite the important information that they can give, as it was recognized very early \cite{GR}. In particular, high frequency measurements can give direct measures of several vortex parameters, such as the pinning constant (or the depinning frequency) and the  free-flux-flow resistivity (see below, Sec.\ref{sec:muw}).

In this paper, we present a preliminary investigation of the pinning properties of S/F/S trilayers (and pure Nb as a comparison) using swept-frequency microwave measurements in the frequency range 1-20 GHz. Since defects in the material can affect severely the vortex pinning, we also performed measurements of the local structure of Nb in our Nb/PdNi/Nb trilayers in order to unveil possible correlations between pinning properties and local disorder. To this aim we exploited the Nb-K edge XAFS technique.

The paper is organized as follows: in Section \ref{sec:exafs} we shortly describe sample preparation and we report measurements of the local structure around Nb by means of XAFS. In Section \ref{sec:muw} we report and discuss our measurements of the vortex parameters obtained from swept-frequency microwave response. In Section \ref{sec:conc} we summarize the results.

\section{Samples and local structure measurements.}
\label{sec:exafs}
Pure Nb films and Nb/Pd$_{84}$Ni$_{16}$/Nb S/F/S trilayers have been grown by ultra-high-vacuum, dc magnetron sputtering. All samples were grown at room temperature. Extensive details on sample preparations have been given elsewhere \cite{IlyinaPhC10}. Single-crystal sapphire substrates were used, in order to avoid the complications arising with the analysis of the microwave response when Si substrates are employed \cite{pompeoSUST05,pompeoJS06}. Three heterostructures were studied. They consisted of two Nb layers of nominal thickness $d_S$= 15 nm, and an inner PdNi layer with thicknesses $d_F$=1, 2 and 8 nm. A pure Nb sample of nominal thickness $d_{Nb}$=30 nm has been examined for comparison. The thickness of the samples has been carefully calibrated against the deposition rate. As expected, the critical temperature $T_c$ dropped with increasing $d_F$ \cite{cirilloPRB09}, and we obtained $T_c$= 7.5, 6.2, 5.1, 4.1 K in the samples with $d_F$= 0 (pure Nb), 1, 2, 8 nm, respectively.

In this paper we are mostly interested in the possible effects of the ferromagnetic thickness on the vortex state properties, which in turn might be sensitive to the structural disorder. Local disorder can affect the superconducting properties in several ways: it can act as a pinning center, thus giving rise to e.g. large critical currents; it can weaken the stiffness of the superfluid, thus giving rise to a smaller superfluid density; it can increase the quasiparticle scattering rate, thus affecting the normal state resistivity and possibly the flux-flow resistivity. Thus, an evaluation of the local disorder is certainly an essential information. To this aim we exploited the XAFS (x-ray absorption fine structure) technique to probe the local atomic structure in the pure Nb film and in heterostructures with $d_F$=2 and 8 nm .

Nb-K edge ($E_{Nb}$=18.986 keV) x-ray absorption spectra (XAS) were collected at the GILDA-BM08 beamline \cite{Gilda} at the European Synchrotron Radiation Facility (ESRF Grenoble, France) in fluorescence geometry. GILDA beamline optics employs double independent crystals Si[311] monochromator and a couple of Pd-coated mirrors for harmonic rejection. Focused x-ray spot (approximately $0.2$mm (v) $\times$ 1 mm(h)) is obtained by vertically bent mirrors and sagittally focused second monochromator crystal. The x-ray absorption spectra were collected in fluorescence geometry keeping the samples at 77 K to reduce the structural disorder induced by thermal vibrations. The sample surfaces were oriented about $45^\circ$ from the polarization direction of the incoming x-ray beam in order to reduce the polarization effects in the spectra.

Nb $k_\alpha$ fluorescence intensity ($E_{K_\alpha}\approx$ 16.6 keV) was measured using a 13-element ultrapure Ge multidetector. Bragg reflections were identified by  comparing the spectra collected with the different elements ($I_i$) and removed by interpolating the data under the peaks; the total fluorescence signal is calculated as $I_f = \sum_i I_i$. The incident  intensity $I_0$ was measured using an Ar filled ionization chamber to calculate $\alpha(E) = {I_f}/{I_0}$, proportional to the Nb linear absorption coefficient  $\mu(E)$ (the thin Nb films ensure negligible self absorption effects). Up to 8 spectra were measured for each sample and averaged up to improve the data statistics. The average statistical noise on the
data, calculated as described in \cite{Estra}, results in the range $1.5-1.8\cdot 10^{-3}$ (Table 1).

\begin{figure}[h]
\centerline{\includegraphics[width=4cm]{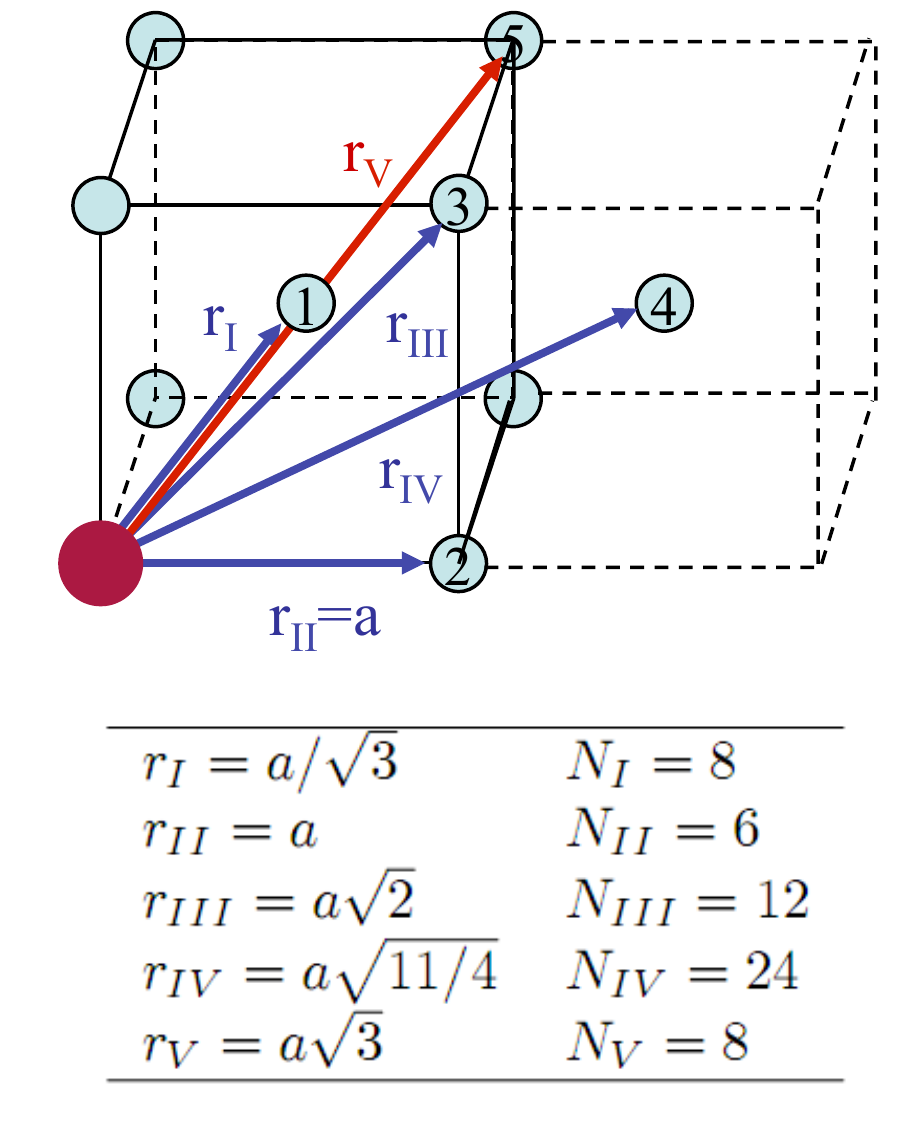}}
\caption{
Local geometry in bcc structure: the five coordination shells used in the analysis are highlighted. The bcc geometry constrains the scattering path lengths $r_i$ and multiplicity numbers $N_i$ of the coordination shells used for XAFS data refinement as summarized in the table.}
\label{fig:bcc}
\end{figure}

The ESTRA and FitEXA codes \cite{Estra} were used for XAS treatment and XAFS quantitative analysis: the $\alpha(E)$ signals were treated using standard procedures for background subtraction, normalization and extraction of  x-ray absorption fine structure (XAFS) signal $\chi^{exp}(k)$~\cite{Estra} ($k =\hbar^{-1}\sqrt{2m_e(E-E_{Nb})}$ is the photoelectron wavenumber).

Quantitative XAFS data analysis has been performed by least square data refinement procedure in the reciprocal space, by fitting the raw $k$-weighted spectra $k\chi^{exp}$ to the theoretical XAFS formula \cite{Lee75}:
\begin{equation}
\label{eq:xafs}
k\chi^{th}(k) = S_o^2\sum_j \frac{N_j\; A_{oj}}{k\;r^2_j}\;\sin(2kr_j+ \phi_{oj}) e^{-\lambda_{oj}/2r_j}\;e^{-2k^2\sigma^2_j}
\end{equation}
Eq.(\ref{eq:xafs}) is a sum of partial contributions coming from single and multiple scattering terms which are calculated assuming a Gaussian model for the pair (or multiple scattering) distribution functions. $N_j$, $r_j$ and $\sigma^2_j$ are the multiplicity (coordination) numbers, interatomic distances and mean square relative displacement (MSRD) of the atomic pair distribution. The $S_o^2$ is an empirical parameter accounting for many body losses, and it was fixed to 0.8 for all the samples. The $A_{oj}$ and $\phi_{oj}$ (amplitude and scattering functions) and $\lambda_{oj}$ (the photoelectron mean free path) were calculated using the FEFF 8.2 code \cite{feff}.

\begin{figure}[h]
 \centerline{\includegraphics[width=5.5cm]{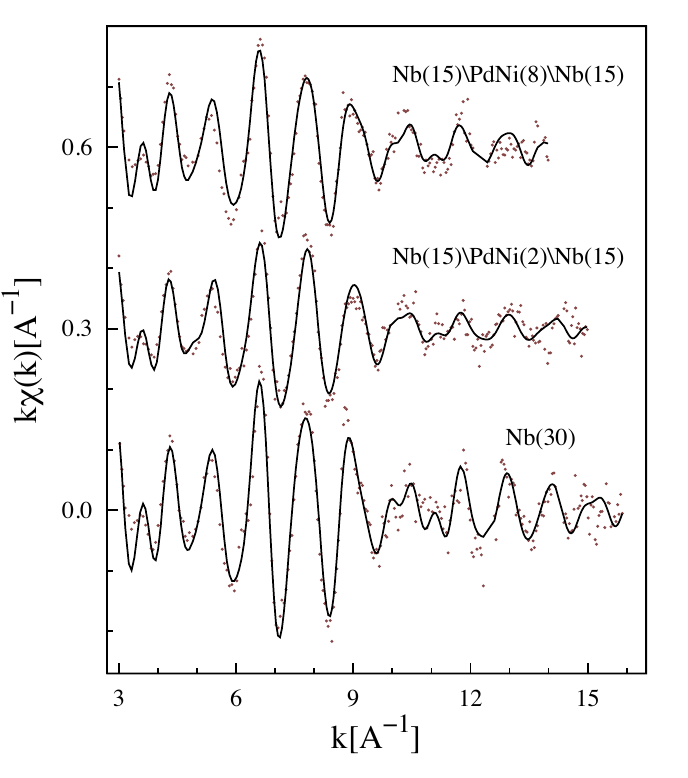}}
  \caption{Experimental Nb K edge XAFS data (points) and best fit (full line) for all the investigated samples (vertically shifted for the sake of clarity).}
\label{fig:exafs}
\end{figure}

\begin{figure}[h]
\centerline{\includegraphics[width=5.5cm]{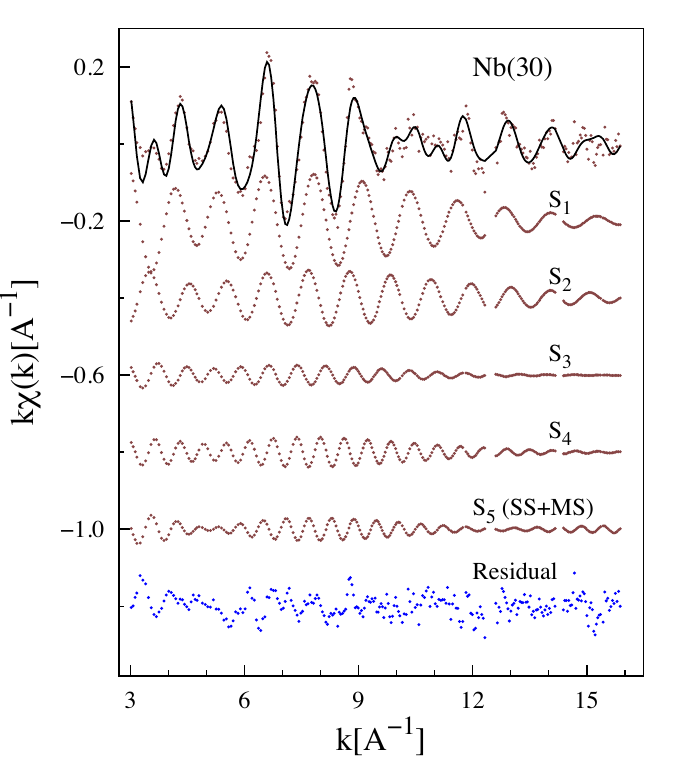}}
  \caption{Example of XAFS data analysis, Nb(30) sample: experimental $k\chi^{exp}(k)$ XAFS data (points) and best fit (full line) are shown at the top. The partial contributions (coordination shells) are shown below, vertically shifted for sake of clarity, labeled according to the bcc geometry (Fig.\ref{fig:bcc}). The residual (experimental minus best fit) is shown at the bottom.}
\label{fig:fit}
\end{figure}

Nb has body centered cubic (bcc) crystallographic structure, depicted in Fig.\ref{fig:bcc}, with lattice parameter (cube edge) a=3.30\AA.
The five coordination shells shown in Fig. 1 were  used to refine the Nb K edge XAFS data. Notice that the aligned Nb$_0-$Nb$_I-$Nb$_V$ configurations along the bcc cube diagonal give rise to strong multiple scattering contributions which have been considered in the analysis. The multiple shell data refinement has demonstrated to improve the XAFS data analysis reliability \cite{MS1,MS2}, as it allows to apply severe constraints among the structural parameters.
In order to reduce the number of free parameters and the correlations among them all the $r_{I}$ distances were constrained to the lattice parameter $a$ (Fig.1) and the multiplicity numbers were fixed to their crystallographic values. The MSRD and the energy scale shift ($\Delta E$) were refined.
An example of best fit, showing the partial contributions and the residual, is reported in Fig.\ref{fig:fit}. The experimental data and best fit for all the three samples investigated are also shown in Fig.\ref{fig:exafs}.
The squared residual function:
$R^2$\cite{Estra} is reported (Table 1) to evaluate the best fit quality.

\begin{table}
  \centering
  \begin{tabular}{ccccccc}
    \hline
    Sample & $R_I$  & $\sigma^2_I$         & $\sigma^2_2$         & $\Delta E$ & $\sigma^2_\chi$ & $R^2$  \\
    & [\AA ] & [$\times$ \AA$^{2}$ ]& [$\times$ \AA$^{2}$ ]& [eV]       & [$\times 10^{-3}$ ]  &  \\
    \hline
    Nb(30)            & 2.85(1)& 5.4(4) & 5.8(6) & -3.9 & 1.8 & 0.13\\
    Nb(15)PdNi(2)Nb(15) & 2.84(1)& 7.7(6) & 8.7(7) & -4.9 & 1.7 & 0.11\\
    Nb(15)PdNi(8)Nb(15) & 2.85(1)& 7.3(6) & 6.9(7) & -4.2 & 1.6 & 0.12\\
    \hline
  \end{tabular}
  \caption{Best fit results of Nb K edge XAFS data analysis.}\label{t.1}
\end{table}

The best fit parameters provided quantitative details about the Nb local structure (Table 1): firstly we notice that the geometrical constraints based on bcc structure provide a satisfactory best fit in Nb(30) as well as in Nb(15)/PdNi(x)/Nb(15) trilayers, thus confirming the validity of the model. The lattice parameter corresponds to what expected in bulk Nb. Therefore, the Nb local structure is close to the bulk Nb structure in all samples, and only weakly affected by the magnetic PdNi layer. Nevertheless, the MSRD (and thus the structural disorder) is larger in trilayers (about 40\%\ higher) than in pure Nb. The growth of the interlayer has then some effect on disordering the Nb atomic structure at the local scale. Interestingly, this effect is nearly independent from the interlayer thickness, as Nb(15)/PdNi(8)/Nb(15) exhibits comparable, or even smaller, disorder with respect to Nb(15)/PdNi(2)/Nb(15).

\section{Microwave measurements.}
\label{sec:muw}

We measured the real part of the complex microwave resistivity in the wide frequency range 1-20 GHz, by means of the so-called Corbino disk technique. The method has been extensively described elsewhere \cite{silvaSUST11}. In short, the thin film terminates a coaxial line. From measurements of the complex reflection coefficient of the line it is possibile, after suitable calibration, to obtain the frequency dependence of the real part of the resistivity. Measurements are performed at fixed field $H$ and temperature $T$, and the frequency is swept up to 30 GHz. The noise increases significantly above 20 GHz, so that in our calibration setup the measurements are reliable only up to $\sim$ 20 GHz. We present here a set of homogeneous measurements, taken in samples with $d_F$ = 1, 2, 8 nm, and in the pure Nb film, at the same reduced temperature $t= T/T_c\simeq$0.86 and at the same reduced field $h=H/H_{c2}\simeq$0.5.

The frequency dependent resistivity allows for an accurate determination of the vortex parameters. On one side, the microwave field induces very short-range oscillations in the vortex system. The mean-square displacement of a vortex line from its pinning site depends on the microwave frequency and field strength, but it is typically much less than 1 nm \cite{tomaschPRB88}. This peculiar feature makes the analysis of the data quite straightforward. In particular, the vortex dynamic can be assumed to be ``single-vortex-like": since the displacement of each vortex is much smaller than the intervortex separation, which with the flux density here applied (between 0.125 T and 0.25 T) is in the $\sim$100 nm range, no perturbation of the static configuration is determined by the microwave current. In this case, the generalized model for vortex motion \cite{pompeoPRB08}, which recovers many well-known models \cite{GR,CC,brandtPRL91} in the appropriate limits, can be applied. The vortex motion complex resistivity reads:
\begin{equation}
\label{eq:rhovm}
\rho_{vm}=\rho_{vm,1}+\rmi\rho_{vm,2}=\rho_{ff}\frac{\epsilon_{eff}+\rmi\nu/\nu_{0}}{1+\rmi\nu/\nu_{0}}
\end{equation}
\noindent where the flux-flow resistivity $\rho_{ff}$ has been discussed previously \cite{torokhtiiPhC12}, $\nu_0$ is a characteristic frequency separating the pinned regime ($\nu<\nu_0$) from the flux-flow regime ($\nu>\nu_0$), and the creep factor $\epsilon\leq1$ takes into account thermally activated processes. The so-called depinning frequency $\nu_p$, which is a commonly used parameter to quantify the strength of vortex pinning, can be easily derived from $\nu_0$ and $\epsilon$, once a specific model is assumed to obtain the analytical expression for $\epsilon$. All the details have been discussed thoroughly elsewhere \cite{pompeoPRB08}.

It is worth mentioning that all three vortex parameters can be measured independently even from the real part of the frequency-dependent vortex motion resistivity. Typically, curves of $\rho_{vm,1}(\nu)$ are fitted to Eq.(\ref{eq:rhovm}), yielding $\rho_{ff}, \nu_0, \epsilon$. Here, in order to derive $\nu_p$, we adopted the Coffey-Clem model \cite{CC}, but we emphasize that different models, such as \cite{brandtPRL91}, gave basically the same results within a scale factor.

In most cases \cite{golosovskySUST96} the microwave complex resistivity has been obtained at a single frequency, typically with the help of some kind of resonating technique, and the analysis of the data has relied on the assumption of vanishing flux creep, which we show below is not the case in our measurements. Thus, in the present case, the frequency-dependent measurements are invaluable in that they allow for the simultaneous measurements of all three vortex parameters.

\begin{figure}[h]
\centerline{\includegraphics[width=5.5cm]{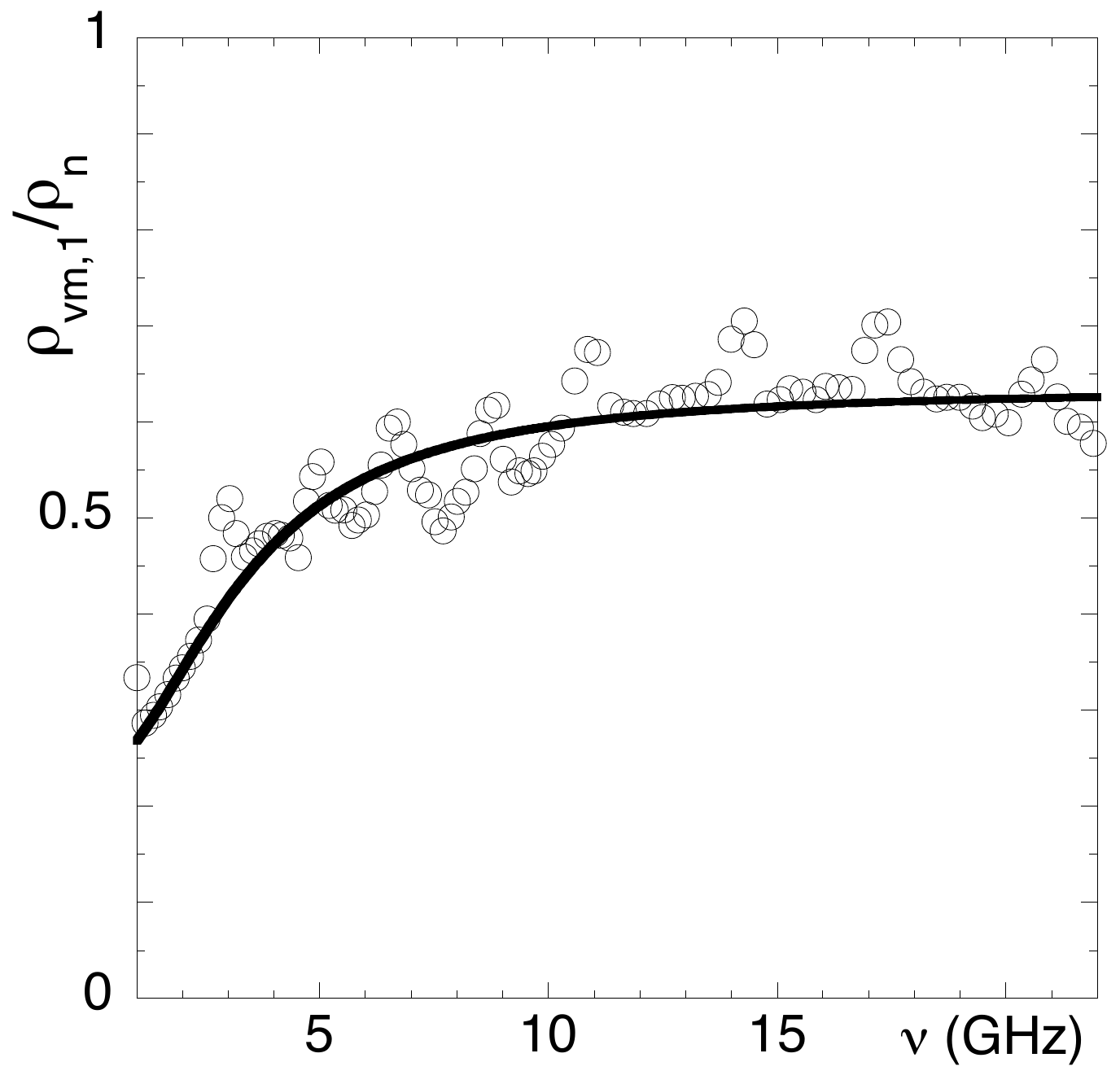}}
  \caption{Full dots: normalized vortex motion resistivity, $\rho_{vm,1}/\rho_n$ vs frequency; continuous line: fit with Eq.(\ref{eq:rhovm}). To avoid crowding, only 15\% of the data are shown. Oscillations are typical of sudden increase of dissipation, that has been discussed elsewhere \cite{pompeoPhC10}. Sample is Nb/PdNi(2nm)/Nb. $T$=4.4 K, $\mu_0H$= 0.2 T.}
\label{fig:data}
\end{figure}

\begin{figure}[h]
\centerline{\includegraphics[width=5.5cm]{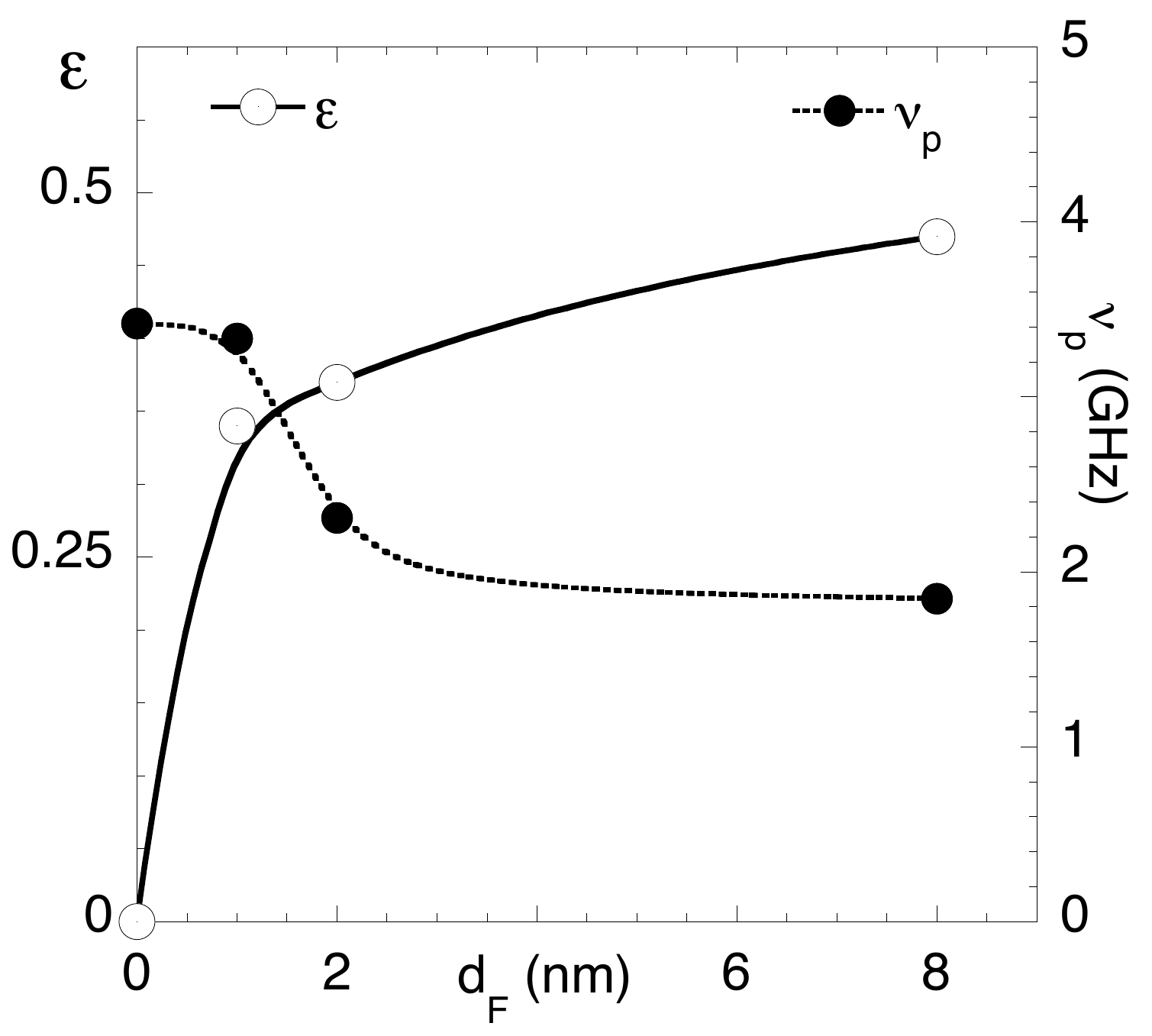}}
  \caption{Pinning parameter $\nu_p$ (depinning frequency), right axis, and creep parameter $\epsilon$, left axis, as a function of the ferromagnetic layer thickness $d_F$. As it can be seen, the increase of the ferromagnetic thickness determines a  decrease of pinning strength and an increase of thermal creep. All measurements have been taken at $T/T_c\simeq$0.86 and $H/H_{c2}\simeq$0.5. Lines are guide to the eye.}
\label{fig:param}
\end{figure}

In Fig.\ref{fig:data} we report typical swept-frequency data taken in one of our samples, together with the fit obtained from Eq.(\ref{eq:rhovm}). Apart from the oscillations, which have been described in \cite{pompeoPhC10}, the overall fit is quite satisfactory and yields the vortex parameters with a small uncertainty.
The vortex parameters are reported in Fig.\ref{fig:param}: here we observe that the ferromagnetic thickness $d_F$ has a dramatic influence on pinning and creep. In pure Nb the depinning frequency $\nu_p\approx$3.5 GHz points to rather strong pinning, in agreement with other estimates \cite{JanjusevicPRB06}. The sample with thinner $d_F$ does not change much in pinning, but the sample with $d_F$=2 nm shows an abrubt collapse of the depinning frequency to $\nu_p\approx$ 2 GHz. By contrast, thermal activation increases sharply immediately when an F layer exists, and steadily keeps increasing with increasing $d_F$. Thus, S/F/S multilayers clearly show a much weaker pinning and larger thermal activation.

By comparing the results of the present Section with the local structure measurements of Sec.\ref{sec:exafs}, we come to the interesting conclusion that if a correlation between the disorder and the change in $\nu_p$ and $\epsilon$ exists, it is far from being trivial. In fact, while the local structural disorder is larger in S/F/S trilayers than in the pure Nb film, the trend is not a monotonous function of $d_F$. The larger disorder is found by XAFS in the sample at $d_F$=2 nm, which does not exhibit neither the weakest pinning, nor the larger thermal activation. Thus, we must argue that the magnetic properties of the F layer play a role in modifying the superconducting response of trilayers.

In fact, both the depinning frequency and the creep factor are related to the condensation energy: the larger the condensation energy of the condensate, the larger the pinning strength, the smaller the creep factor. Thus, it can be reasonably argued that the F layer affects vortex pinning and creep through a reduction of the condensation energy. This explanation would be in agreement with the observed reduction of the superfluid fraction observed in S/F bilayers \cite{lembergerJAP08}.

\section{Summary.}
\label{sec:conc}

We have presented measurements of the frequency-dependent vortex motion resistivity in pure Nb and Nb/PdNi/Nb trilayers. We have shown that vortex pinning and vortex creep are heavily affected by the presence and thickness of the F layer: pinning decreases and creep increases with increasing F thickness. Details about the local atomic structure in S phase (Nb) obtained from XAFS analysis demonstrate that structural disorder has an effect in modifying the pinning and creep strength, however the structural disorder alone cannot explain the observed weakening of pinning/strengthening of creep as a function of F-layer thickness. Combined XAFS and microwave results are relevant to recognize the role of F layer in modifying the superconducting state itself.
While deeper investigations are necessary to  clarify fully the effect of the F layer on the superconducting properties of the heterostructures, we argue that the effects observed in the vortex response might arise from a reduction of the condensation energy.
\begin{acknowledgements}
This work has been partially supported by an Italian MIUR-PRIN 2007. GILDA project is financed by Italian institutions: CNR, INFN.
\end{acknowledgements}

\end{document}